\documentstyle[prl,aps,graphicx,epsf]{revtex}
\begin{document}
\title{In-Plane Elliptic Flow of Resonance Particles\\
in Relativistic Heavy-Ion Collisions}
\draft
\author{Tetsufumi Hirano}
\address{Department of Physics, Waseda University, Tokyo 169-8555, Japan}
\date{\today}
\maketitle
\begin{abstract}
			
We analyze the second Fourier coefficient $v_2$ of the pion azimuthal
distribution in non-central heavy-ion collisions in a relativistic
hydrodynamic model.
The exact treatment of the decay kinematics of resonances
leads to almost vanishing azimuthal anisotropy of pions near
the midrapidity, while the matter elliptic flow is in-plane at freeze-out.
In addition, we reproduce the rapidity dependence of $v_2$
for pions measured in non-central Pb + Pb collisions at 158$A$ GeV.
This suggests that resonance particles as well as stable particles
constitute the in-plane flow and are important ingredients
for the understanding of the observed pion flow. 
\end{abstract}

The main goals of relativistic nuclear collisions are to
determine the nuclear equation of state (EOS) under extreme
conditions and to understand a new phase of deconfined nuclear matter,
the quark-gluon plasma (QGP) \cite{QM99}. 
Since it is the pressure gradient perpendicular to the collision axis
that causes various transverse collective flows, such
as radial flow, directed flow, and elliptic flow, in relativistic
nuclear collisions, 
these flows observed in the final state 
are expected to carry the information about the
EOS\cite{OLLIQM97}.
If the QGP phase is created in nuclear collisions, the quark matter
expands, cools down, and goes through the ``softest point" \cite{HUNG95}
where the ratio of the pressure to the energy density takes its minimum
as a function of the energy density. Therefore, it is expected that
the suppression of the collective flows is not only a signal for
the existence of the QGP \cite{HUNG95,OLLI92,RISCH1996,SORGE99} but also
a useful tool to determine the EOS near the phase transition region.

Some experimental groups reported that the inverse slope parameters
of the transverse mass spectra for the non-multistrange hadrons
$\pi$, $K$, $p$, and $d$ in central Pb + Pb collisions at 158$A$ GeV at
the CERN Super Proton Synchrotron (SPS) are parametrized by two
common values, the freeze-out temperature and the transverse flow
velocity \cite{1997NA44}.
This implies that these particles constitute the radial
flow and the local thermalization at least among those particles
is achieved in {\it central} collisions. It has been, however,
an open question whether equilibration is achieved even
in {\it non-central} collisions. 
In this Letter, we study the elliptic flow that is made of
stable particles and resonances with a hydrodynamic model,
and show that the final pion distribution is well described
in this approach,
if decay kinematics is appropriately taken into account.

The rapidity dependence of azimuthal anisotropy for particle $i$ is
characterized by the coefficients $v_n^i$ $(n=1, 2, 3...)$
in the Fourier expansion of the azimuthal distribution of
the particle \cite{POSK99}:
\begin{eqnarray}
\frac{dN^i}{p_T dp_T dY d\phi}
= \frac{1}{2 \pi}\frac{dN^i}{p_T dp_T dY}[1 + 2 v_1^i (Y, p_T) \cos \phi + 2 v_2^i (Y, p_T) \cos 2 \phi + \cdots ],
\end{eqnarray}
where $\phi$ is the azimuthal angle measured from the
reaction plane and $Y$ is the rapidity.
Non-vanishing $v_1$ and $v_2$ imply
the formation of directed and elliptic flows, respectively.
Experimentally, $v_2 (Y)$ is of the order of four percent around
the midrapidity ($Y=2.92$) for low transverse momentum
($50 < p_T < $ 350 MeV/$c$) charged 
pions in non-central Pb + Pb collisions
that correspond to the impact parameter range
$6.5 < b < 8.0$ fm \cite{1998NA49}. The value of the observed $v_2$
has been claimed to be smaller than that predicted for {\it direct}
pions in hydrodynamic models. 

In the following, we consider not only direct pions but also indirect pions
that are from resonance decays. Thus, we need to strictly distinguish
the matter flow before freeze-out and the observed particle flow.
In this Letter
we use a term ``in-plane" when the hydrodynamic flow before
freeze-out is directed preferentially to the positive and the negative
$x$-axes on the transverse plane and a term ``positive
elliptic flow" when $v_2$ of observed particles
is positive. 
Here the $x$-axis is defined as the direction of impact parameter
in non-central collisions.
When one only considers particles directly emitted from freeze-out
hyper-surface, ``positive elliptic flow" means that the hydrodynamic
flow is ``in-plane". However, once
feeding from resonance decays is included,
the above two are no longer equivalent.

First, let us suppose that a resonance particle with  mass $m_{R}$
is emitted from freeze-out hyper-surface and decays
into two identical daughter particles with mass $m$ in the vacuum.
In the rest frame of the resonance particle the decay is isotropic
and the daughter particles are uniformly distributed with respect to
the azimuthal angle when one averages the spin of the resonance
particles.
This is, however, not the case anymore if the resonance is moving in
a reference frame. To see this more clearly,
let us assume that the resonance particle moves with velocity
$(V_{Rx}, V_{Ry}, V_{Rz}) = (V, 0, 0)$ in a reference frame.
If $V$ is larger than the critical value 
$V^* = p^*/E^*$, where
$p^* = \sqrt{m_{R}^2/4-m^2}$ and $E^* = m_{R}/2$, 
the probability that the daughter particle is emitted with an angle
$\phi$ from the $x$-axis peaks at 
$\phi_{\pm} = \pm \sin ^{-1} [ p^* (1-V^2)^{1/2}/(m V) ]$
and $\phi$ is limited in a range,
$\phi_{-} \le \phi \le \phi_{+}$ \cite{LANDAU}.
The two peaks are due to the Jacobian singularity in the
Lorentz transformation from the resonance
rest frame to the reference frame.
As a result, 
the opening angle between 
the daughter particles tends to remain finite
even if the resonance particle moves
at a large velocity. This is the reason why the
equivalence between ``in-plane flow" and ``positive elliptic flow"
breaks down when one includes the decay of resonances in the final state.

\begin{figure}
\begin{center}
\includegraphics[width=12cm]{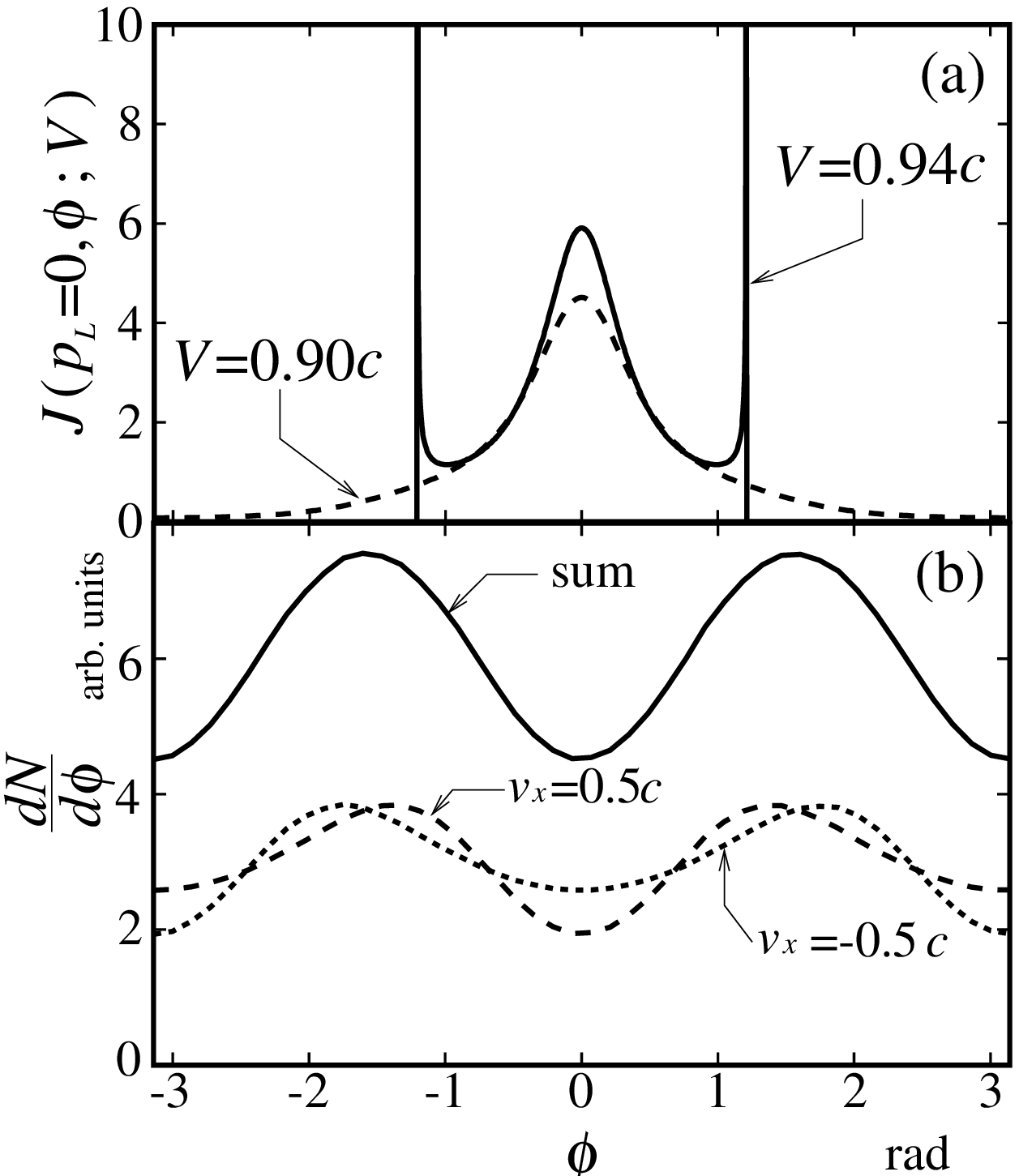}
 \caption{(a)Jacobian as a function of azimuthal angle
in the Lorentz transformation between two reference frames (see text).
The solid and dashed lines
correspond to $V_{\rho} = 0.94c$ and $0.9c$, respectively.
(b)Azimuthal distribution (in arbitrary units) of pions at
midrapidity through decays of $\rho$-mesons from two
fluid elements which are moving in opposite directions.
The velocities of fluid elements are assumed to be
$(v_x, v_y, v_z) = (\pm 0.5c$, 0, 0).}
 \label{FIG1}
\end{center}
 \end{figure}

The multiplicity of pions through two body decays of
resonances is given by
\begin{eqnarray}
\label{MULTI}
N_{R \rightarrow \pi X} & = &
\int \frac{J(p_L,\phi;{\bf V}_{R})dp_L d\phi}{4\pi p^*}
B_{R \rightarrow \pi X} \frac{d^3 {\bf p}_{R}}{E_{R}} \int ds W_{R}(s) \frac{d_R}{(2 \pi)^3}
\frac{p_{R}^\mu d\sigma_\mu}{\exp[(p_{R}^\nu u_\nu - \mu)/T_{f}] \mp 1},
\end{eqnarray}
where $u^\nu$, $\mu$,
$T_f$, and $\sigma_\mu$ are, respectively,
the four-dimensional fluid velocity,
the chemical potential,
the freeze-out temperature, and
the freeze-out hyper-surface.
$p_L$ is the longitudinal momentum of pion. 
$p_R^\nu$ and $d_R$ are, respectively, 
the resonance four momentum in the reference frame and the degeneracy,
and $-(+)$ is for boson (fermion) resonances. 
$B_{R \rightarrow \pi X}$ is the branching ratio of
the decay process and $W_{R}$
is the Breit-Wigner type function, which takes account 
of the finiteness of the resonance width. For $W_{R}$, we adopt the
form used in Ref.~\cite{SOLL91}. 
The Jacobian of the Lorentz transformation from the resonance
rest frame to an arbitrary reference frame $J(p_L,\phi;{\bf V}_{R})$ is defined by
$dp_L^* d\phi^* = J(p_L,\phi;{\bf V}_{R})dp_L d\phi$,
where the quantities with (without) $*$ are the ones in the resonance
rest (reference) frame.
Two typical shapes of the Jacobian as functions of
the azimuthal angle in $\rho \rightarrow \pi\pi$ are shown
in Fig.\ \ref{FIG1} (a). If ${\bf V}_\rho = (V_\rho, 0, 0)$
and $V_\rho < V^* (\sim 0.93 c 
{\rm ~in~this~process})$,
$J(p_L=0,\phi; {\bf V}_\rho)$ has a broad peak around $\phi = 0$
at the same azimuthal angle
of the resonance particle as expected. However, if $V_{\rho} > V^*$,
$J(p_L=0,\phi; {\bf V}_\rho)$ has a finite value only in the range
$\phi_{-} < \phi < \phi_{+}$ and two sharp peaks
appear at $\phi_{\pm}$ in addition to the original broad one
as explained above.

In order to get an idea about the effect
of thermal smearing, we estimate the azimuthal distribution of
pions through the above process, taking the following simple
model: There are only two fluid elements with local
fluid velocities $(v_x, v_y, v_z) = (\pm 0.5c, 0, 0)$.
Both elements are assumed to be thermalized at 
the temperature $T_{f}=120$ MeV. The pion distribution from
$\rho$-meson decays in each fluid element is shown in Fig.\ \ref{FIG1}
(b).
Here we sum up the pion distribution over $50 < p_T < 350$ MeV and $-0.5 < Y < 0.5$.
The sharp peaks at $\phi \sim\pm 1.21$ ($\pm 1.93$) in the
Jacobian are smeared by thermal motion of $\rho$-mesons, but
they are still visible in the
azimuthal distribution of pions from $\rho$-mesons in the fluid element with
$v_x = 0.5c$ ($-0.5c$),
while the peaks at $\phi = 0$ ($\pi$)
are completely washed out.
The superposition of the two distributions
leads to an azimuthal distribution with broad peaks at $\pm \pi/2$,
as shown in Fig.\ \ref{FIG1} (b). 
In this simple model,
pions from $\rho$-meson decay have negative $v_2$, i.e.,
the elliptic flow in the final state is negative,
while the motion of the two fluid elements is in-plane.

We next carry out realistic hydrodynamic simulations
for the space-time evolution of Pb + Pb collisions at
158$A$ GeV to see how large the effect of the Jacobian singularity
is for the final state pion distribution.
We first assume
that the hot and dense nuclear matter produced in heavy-ion
collisions is in local thermal equilibrium after $t_0 = 1.44$ fm/$c$
since the two nuclei touched \cite{INITIALTIME}.
Then, we describe the space-time evolution of nuclear matter
after this time by using a relativistic hydrodynamic model without
assuming the Bjorken's scaling solution\cite{BJOR} or the cylindrical
symmetry along the collision axis \cite{AMELIN,95RISCH,NONAKA,HIRANO}.
Thus, it is possible to discuss 
the rapidity dependence of elliptic flow $v_2(Y)$ for charged pions
through resonance decays as well as for charged pions directly emitted
from the freeze-out hyper-surface in this model.
We use a model EOS with a first order phase transition
between the QGP phase and the hadron phase \cite{NONAKA}.
The QGP phase is assumed to be free gas composed of quarks with
$N_f = 3$ and gluons. For the hadron phase we adopt a resonance
gas model, which includes all baryons and mesons up to
the mass of 2 GeV \cite{PDG}, together with an excluded volume
correction \cite{RISCH93}.
We use the critical temperature at zero baryon density,
$T_{c}(n_{B}=0) = 160$ MeV.
The two model EOS's are
matched by imposing the Gibbs' condition for phase equilibrium
on the phase boundary.
The numerical results of the hydrodynamic simulation give us
the momentum distribution of hadrons through the Cooper-Frye
formula \cite{CF} with a freeze-out energy density $E_{\rm f} = 60$ MeV/fm$^3$.
We have used this formula for the {\it direct} emission of $\pi^-$,
$K^-$, and $p$.
In addition, we have taken into account negative pions
from the decays of $\rho$, $\omega$, $K^*$,
and $\Delta$ in the final state.

In the numerical simulation we have fixed the impact parameter at 7.2 fm.
We have chosen the initial parameters in the hydrodynamic simulation
to reproduce not only the rapidity and transverse mass distribution of negative hadrons in central collisions \cite{NA49RAP} but also the preliminary data of rapidity distribution of negative pion in non-central collisions \cite{COOPER} by the NA49 collaboration.
The corresponding central energy density $E_0$ and
baryon number density $n_{B0}$ are, respectively, 3.9 GeV/fm$^3$
and 0.46 fm$^{-3}$.
The energy density and the baryon number density
on the transverse plane are assumed to be in proportion to
the number of wounded nucleons with the standard Woods-Saxon distribution for the nuclear density.
We have assumed the initial transverse flow to vanish.
For details on our hydrodynamic model, see Ref.~\cite{HIRANO}.

\begin{figure}
\begin{center}
\includegraphics[width=12cm]{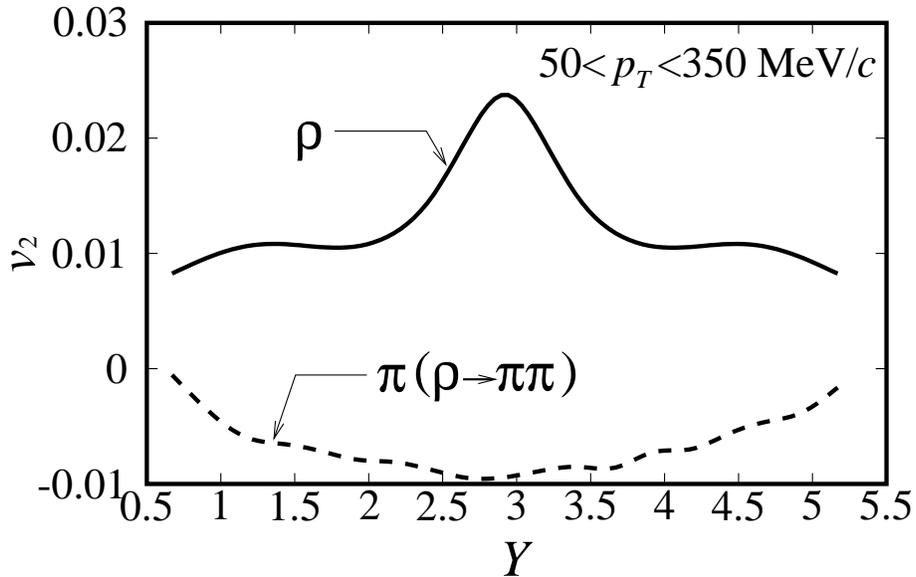}
 \caption{Azimuthal anisotropy $v_2$ for $\rho$-mesons
directly emitted from freeze-out hyper-surface (solid line)
and for pions through $\rho \rightarrow \pi\pi$ (dashed line)
in non-central Pb + Pb collisions at 158 $A$ GeV.}
 \label{FIG2}
\end{center}
 \end{figure}

We first discuss the effect of the Jacobian singularity for
the pions through $\rho$-meson decays in the realistic hydrodynamic
calculation.
The momentum distribution
of $\rho$-mesons is free from the effects of the Jacobi function
and is given by the second integral in Eq.~(\ref{MULTI}),
i.e., the integral with respect to $s$ and $\sigma_\mu$.
In Fig. \ref{FIG2}, we show $v_2$ for the $\rho$-mesons
directly emitted from freeze-out hyper-surface and for the pions through
$\rho \rightarrow \pi\pi$. The elliptic flow of $\rho$-mesons
is positive and in-plane.
Nevertheless, the $v_2$ for the pions through $\rho \rightarrow \pi\pi$
almost vanishes near the midrapidity due to the decay kinematics.
This implies that the effect of the Jacobian singularity, which we discussed
above with a simple model, survives even in this realistic
calculation. The behavior of the $v_2$'s of pions from
$K^*$ or $\Delta$
is similar to this.

\begin{figure}
\begin{center}
\includegraphics[width=12cm]{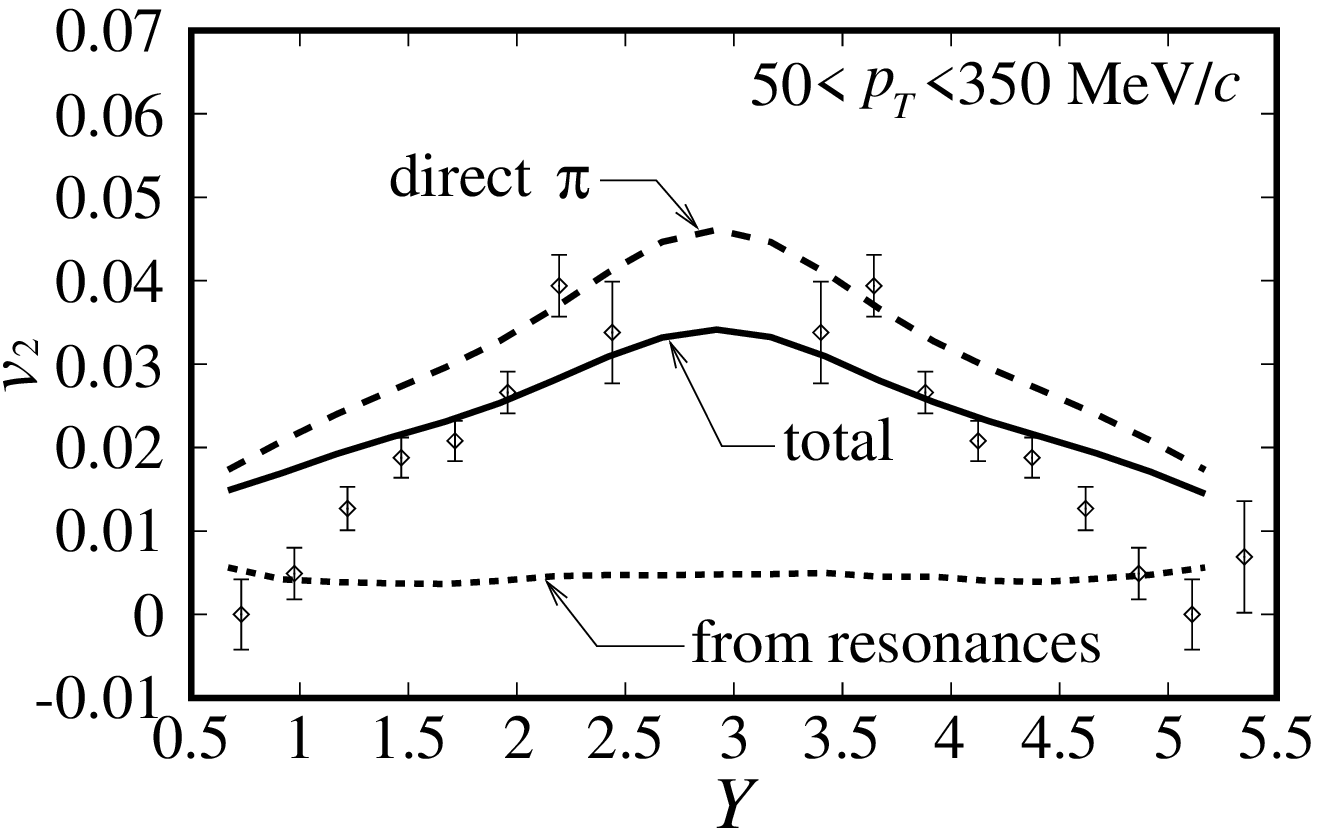}
 \caption{Azimuthal anisotropy $v_2$ for total charged pions (solid line), direct pions (dashed line), and pions from resonance decays (dotted line) as
functions of rapidity. The experimental data was measured by NA49 in Pb + Pb collisions at 158 $A$ GeV.
See text for details.}
 \label{FIG3}
\end{center}
 \end{figure}

Finally, we discuss the rapidity dependence of
the observed pion elliptic flow obtained from our fully
three-dimensional hydrodynamic calculation.
In Fig.\ \ref{FIG3} we compare our results with the experimental
data by the NA49 Collaboration \cite{1998NA49}.
The solid line represents $v_2$ for the total charged pions.
For comparison, $v_2$ for pions directly
emitted from freeze-out hyper-surface (dashed line)
and that for pions through resonance decays (dotted line)
are shown separately.
Our results were obtained by summing up pion distribution
over a $p_T$ range, 
$50 < p_T < 350$ MeV/$c$. The experimental data
corresponds to an impact parameter range $6.5 < b < 8.0$ fm.
The solid line
is in good agreement with the experimental data near midrapidity \cite{HBT}.
The $v_2$ for indirect pions from resonance decays vanishes.
This reduces $v_2$
for the total pions by about 26 \% at midrapidity.
This figure tells us that hydrodynamic description, which assumes
the local thermal equilibrium, works well also for the expansion stage
of non-central collisions at the SPS energy \cite{KOLB2}.

A few remarks on other calculations are in order here.
At midrapidity, our result for the total pions is
consistent with the previous result obtained by a two-dimensional
hydrodynamic model \cite{KOLB2}. The authors of Ref.\ \cite{KOLB2}
assumed that the longitudinal expansion can be described by the
Bjorken's scaling solution \cite{BJOR} and numerically
simulated the evolution of nuclear matter only in the transverse directions.
Hence they could not obtain the rapidity dependence of elliptic flow.
They also took into account resonance particles up to the mass of the
$\Delta$(1232), but concluded that resonance decays reduce
the momentum anisotropy for pions by only 10-15 \%.
Their reduction factor corresponds to full $p_T$ range \cite{KOLB}.
The effect of Jacobian singularity is important in \textit{low} $p_T$ region.
When we integrate the distribution over full $p_T$ range, we obtain the result reduced by 11 \% .
Liu {\it et al.} \cite{RQMD} compared their results from a transport
model, the Relativistic Quantum Molecular Dynamics (RQMD), with
the experimental data and concluded that the model calculations are
in reasonable agreement with experimental data. The experimental
data of elliptic flow was, however, later updated \cite{1998NA49}, and
the agreement is not as good as before anymore.
Soff {\it et al.} \cite{URQMD} also obtained $v_2(Y)$ for pions
from a microscopic transport model, the Ultrarelativistic
Quantum Molecular Dynamics (UrQMD), but the situation is
similar to the RQMD model. 
Recently, it was argued that the suppression of the elliptic flow
is due to partial thermalization \cite{OLLIQM97,HEISEL,VOLO00}.
However, as we have shown, this is not needed to explain
the data. According to our result, the suppression
rather implies the full thermalization
of the system, including resonances.

In summary, we have investigated the 
elliptic flow of pions in Pb + Pb non-central collisions at 158$A$ GeV in 
a relativistic hydrodynamic model. As sources of pions,
we have considered not only direct pion emission from the freeze-out
hyper-surface but also decays of resonance particles.
Pions from resonance decays suppress the azimuthal
anisotropy in the midrapidity region as much as 26 \%.
Taking this effect into account, we were able to reproduce the
experimental data of the rapidity dependence of $v_2$.
These results lead to our conclusion that the pions and
the resonance particles
constitute thermalized in-plane elliptic flow and that the hydrodynamic
picture is applicable to the expansion stage in the non-central
collisions at the SPS energy. The Jacobian singularity effect
should exist also in cascade calculations such as RQMD and UrQMD
if the decay kinematics is appropriately taken into account.
It will be interesting to see how important this mechanism is for
the suppression in those calculations.

The author is much indebted to Professor I. Ohba and Professor
H. Nakazato for their helpful comments. He wishes to acknowledge
valuable discussions with M. Asakawa, T. Hatsuda, P. Huovinen, P. F. Kolb, T. Matsui, Y. Miake, S. Muroya, S. Nishimura, C. Nonaka, A. Ohnishi, J.-Y. Ollitrault, and A. M. Poskanzer.
He also thanks M. Asakawa for a careful
reading of the manuscript and C. Nonaka for giving him a numerical
table of EOS.


%
%
%
%


\begin{references}
\bibitem{INITIALTIME}$t_0$ is the passage time when the Lorentz contracted incident nucleus finishes penetrating the target nucleus in the center of mass system.
\end{references}
\end{document}